\documentclass[aps,prl,twocolumn,showpacs,amsmath,amssymb,superscriptaddress]{revtex4-1}

\usepackage{graphicx}
\usepackage{color}
\usepackage{epstopdf}
\usepackage{amsmath}
\usepackage{bm}

\newcommand{\reffig}[1]{Fig.~\ref{#1}}
\newcommand{\refeq}[1]{Eq.~(\ref{#1})}

\begin{document}

\title{Supersolidity around a critical point in dipolar Bose Einstein condensates}
\author{Yong-Chang Zhang}
\author{Fabian Maucher}
\author{Thomas Pohl}
\affiliation{Department of Physics and Astronomy, Aarhus University, Ny Munkegade 120, DK 8000 Aarhus, Denmark}

\begin{abstract}
We explore spatial symmetry breaking of a dipolar Bose Einstein condensate in the thermodynamic limit and reveal a critical point in the phase diagram at which crystallization occurs via a   second-order phase transition. This behavior is traced back to the significant effects of quantum fluctuations in dipolar condensates, which moreover stabilize a new supersolid phase, namely a regular honeycomb pattern with maximal modulational contrast and near-perfect superfluidity. 
\end{abstract}
\maketitle

The crystallization of a liquid typically proceeds via a first-order phase transition, associated with the release of latent heat whereby fluid and solid phases coexist during the freezing stage. While this scenario applies to a wide range of classical and quantum systems, quantum mechanical effects facilitate another form of coexistence, where superfluidity and crystalline spatial order can simultaneously occur in a so-called supersolid phase. The coexistence of such seemingly exclusive properties was first pointed out by Gross \cite{Gross:PR:1957}, who suggested the possibility of a density-modulated superfluid of weakly interacting Bosons described by a classical meanfield. Subsequent theoretical work \cite{Andreev:JETP:1969,Chester:PRA:1970,Leggett:PRL:1970} has since motivated an extensive search for supersolidity in low-temperature helium experiments \cite{suso_Rev,suso_RMP} over the past several decades.

More recently, dilute ultracold atomic quantum gases have emerged as a promising alternative \cite{CA1,Henkel:PRL:2010,CA2,CA3,CA4} to observe the elusive supersolid phase of matter. Indeed the spontaneous emergence of density modulations in the presence of phase coherence has been realized by applying external light fields to Bose Einstein condensates in optical cavities \cite{Donner:nature:2017} and to implement synthetic spin-orbit coupling \cite{Ketterle:nature:2017}. Experimental breakthroughs in realizing dipolar quantum gases of dysprosium \cite{Dy1} and erbium \cite{Er1} atoms have raised promise for observing supersolidity arising from their atomic interactions. Indeed a series of recent experiments have revealed a roton-maxon excitation spectrum \cite{roton1} in close analogy to the physics of superfluid helium \cite{Landau:PR:1949}, and reported the observation of pattern formation \cite{Pfau:nature:2016} as well as self-confined quantum droplet states \cite{Pfau:nature2:2016,Santos:PRX:2016,Ferrier_Barbut:PRL:2016}, arising from the interplay of short-range collisions, dipolar interactions and quantum fluctuations \cite{Santos:PRA:2016,Santos:PRA2:2016}. Theoretical work has explored the formation of regular quantum droplet arrangements under various conditions \cite{Santos:PRA:2016,Bisset:PRA:2016,Blakie:PRL:2018,1D_1,Tanzi} and the temporary coexistence of phase coherence and density modulations along one spatial dimension \cite{Tanzi,1D_2} has been reported recently. Yet, the preparation of extended supersolid ground states remains challenging, partly due to the expected first-order nature of the crystallization transition and the associated discontinuous increase of the modulation amplitude, which limits the achievable phase coherence of the modulated state and leads to the generation of high-energy excitations \cite{Pfau:nature:2016} upon dynamically crossing the crystallization line.

\begin{figure}[t!]
\centering
  \includegraphics[width=1\columnwidth]{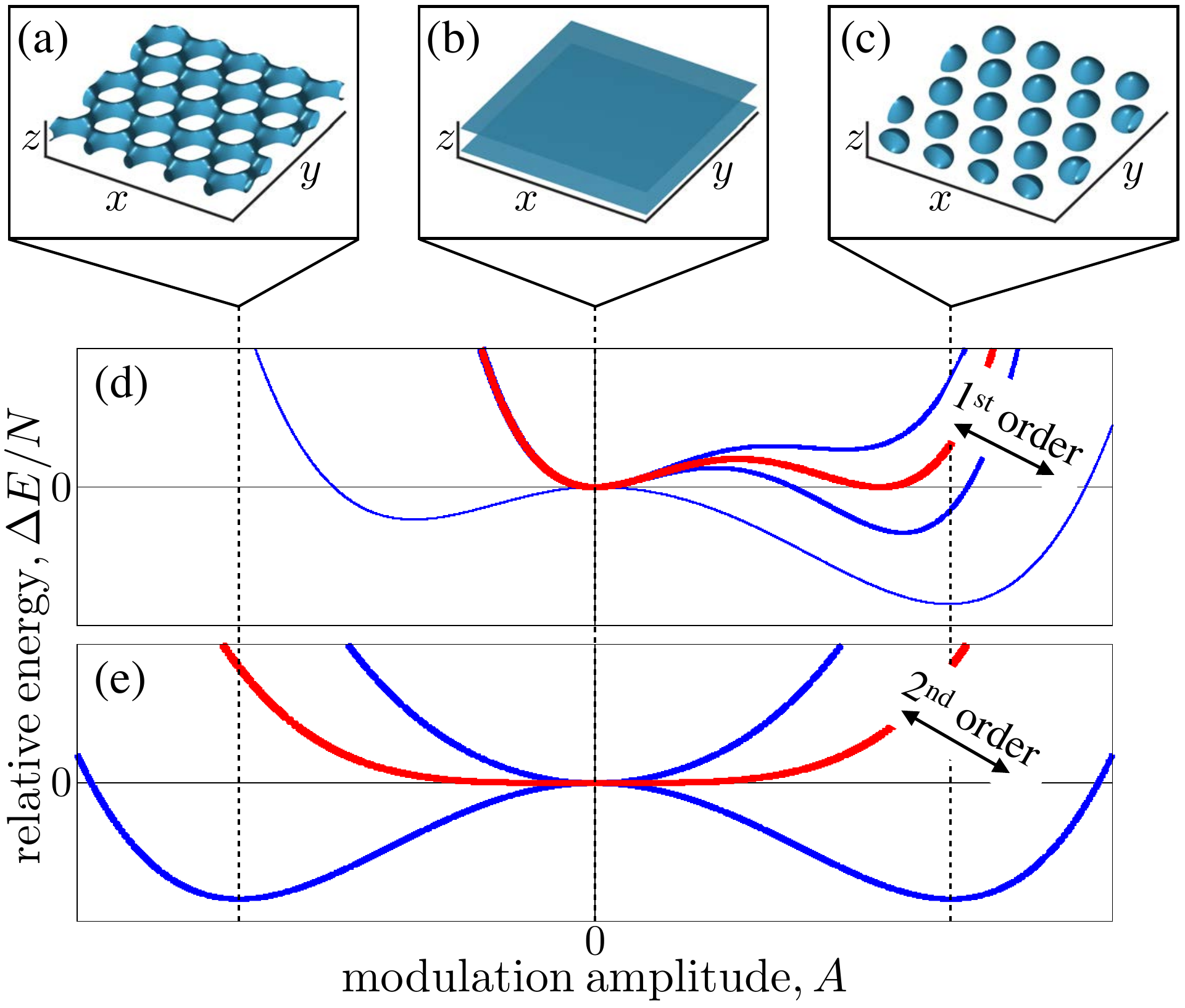}
\caption{
(color online) Iso-density surfaces of the three possible groundstates (a-c) in dipolar Bose Einstein condensates, considered in this work. Apart from an unmodulated superfluid (b), states with a  broken continuous translational symmetry (a,c) exist in the form of a triangular droplet lattice (c) and a honeycomb structure (a). The quantum phase transitions between them can be first-order (d) or second-order (e), as illustrated by the dependence of ground state energy on the modulation amplitude, $A$, of the symmetry-broken state. Panels (d) and (e) show the energy relative to that of the unmodulated superfluid ($A=0$) at (red line) and close (blue lines) to the respective phase transition.
}
\label{fig1}
\end{figure}

Here, we investigate spatial symmetry breaking of dipolar condensates in the thermodynamic limit and show that the crystallization transition can become second-order at a critical point in the underlying phase diagram. In this case, spatial symmetry breaking occurs via the gradual emergence of a two-dimensional density-wave (see Fig.\ref{fig1}), giving rise to an extended supersolid region in the underlying phase diagram. The origin of this behaviour is traced back to the dominant role of quantum fluctuations, which can cancel kinetic energy contributions that would otherwise lead to a first-order transition in typical crystallization scenarios ~\cite{pattern_RMP}. This competition of energy scales also leads to the emergence of a new ordered phase -- a honeycomb structure [see Fig.\ref{fig1}(a)] that is solely stabilized by quantum fluctuations and maintains near-perfect superfluidity.

We consider a zero-temperature quantum gas of $N$ dipolar Bosonic atoms with a mass $m$ which are harmonically trapped along the dipolar polarization axis but otherwise unconfined in the two-dimensional plane perpendicular to it. The particles interact via zero-range collisions with a scattering length $a_{\rm s}$ and via dipolar interactions, characterized by an associated length scale $a_{\rm dd}$. 
The condensate wave function $\psi({\bf r})$ of the atoms is normalized to unity and hence $\rho=N|\psi|^2$ defines the atomic density.
In the limit of weak interactions and upon expressing all spatial coordinates in units of $\ell=12\pi a_{\rm dd}$ as well as scaling time by $m\ell^2/\hbar$ one can express  the total energy $E$ of the condensate as 
\begin{equation}
 \frac{E}{N}=\int \frac{|\nabla\psi|^2}{2} + U(z)|\psi|^2+\frac{2}{5}\gamma N^{3/2}|\psi|^{5}{\rm d}{\bf r}+\frac{E_{\rm I}}{N}. \label{eq:energy}
\end{equation}
The first and second term account for the kinetic energy of the atoms and the axial trapping potential $U(z)=\frac{1}{2}\omega^2_z z^2$, respectively, where $\omega_z$ denotes the dimensionless trap frequency in the units introduced above. The third contribution describes the effects of quantum fluctuations to leading order in the strength of the atomic interactions, as given by the Lee-Huang-Yang (LHY) correction \cite{LHY1,LHY2} , where $\gamma=\frac{4}{3\pi^2}(\frac{a_{\rm s}}{3a_{\rm dd}})^{5/2}[1+\frac{3}{2}(\frac{a_{\rm dd}}{a_{\rm s}})^2]$~\cite{Bisset:PRA:2016}. 
This expression is based on a local density approximation~\cite{Pelster:PRA:2011}, whose accuracy has been demonstrated through comparisons with numerical results from quantum Monte Carlo simulations~\cite{Saito:SocJap:2016}. Being the result of a perturbative expansion, this term should generally be a small correction to the individual mean field interaction of the atoms. Yet, it can have profound effects on the condensate behaviour for specific situations, such as quantum gas mixtures, where interactions of opposite signs can cancel overall mean field effects \cite{Petrov,Tarruell:Science:2018}, or the present case of dipolar Bosons, whose interaction features repulsive  and attractive contributions \cite{Santos:PRA:2016,Blakie:PRA:2016}.  The corresponding mean field interaction energy 
\begin{equation}
E_{\rm I}=N^2\!\!\!\int\!\!\frac{a_{\rm s}}{6a_{\rm dd}}|\psi({\bf r})|^4+\frac{|\psi({\bf r})|^2}{8\pi}\!\!\int\!\! V({\bf r}-{\bf r}^\prime)|\psi({\bf r}^\prime)|^2{\rm d}{\bf r}^\prime {\rm d}{\bf r}\label{eq:Epot}
 \end{equation}
is composed of a contribution from zero-range collisions, proportional to $a_{\rm s}$, and long-range dipole-dipole interactions with the interaction potential $V({\bf r})=(1-3z^2/r^2)/r^3$. 

Without external confinement, the condensate can form a self-confined quantum droplet ground state, which exhibits an elongated shape and continually grows with increasing particle number. The interaction between two such elongated filaments \cite{Fabio:PRL:2017} turns out to decrease upon increasing their length. Indeed one can show that their interaction eventually vanishes in the limit of infinitely extended filaments and therefore precludes the emergence of any discrete translational symmetry. 
A finite trapping potential along the $z$-axis is thus needed to facilitate the formation of finite filaments, which generates long-range interactions between them and thereby enables the formation of extended droplet crystal ground states. 

A reliable characterization of the associated quantum phase transition requires a high numerical accuracy of the ground state and its energy. To this end, we minimize~\refeq{eq:energy} by starting from a modulated state 
whose crystal structure and lattice constant are compatible with the numerical box, on which
we impose periodic boundary conditions perpendicular to the trap axis. 
While the subsequent imaginary-time evolution typically relaxes the system to a local energy minimum defined by the initial state, we find the global ground state by minimizing the obtained energy with respect to the seeded lattice constant and various different underlying structures.

\begin{figure}[t!]
\centering
  \includegraphics[width=\columnwidth]{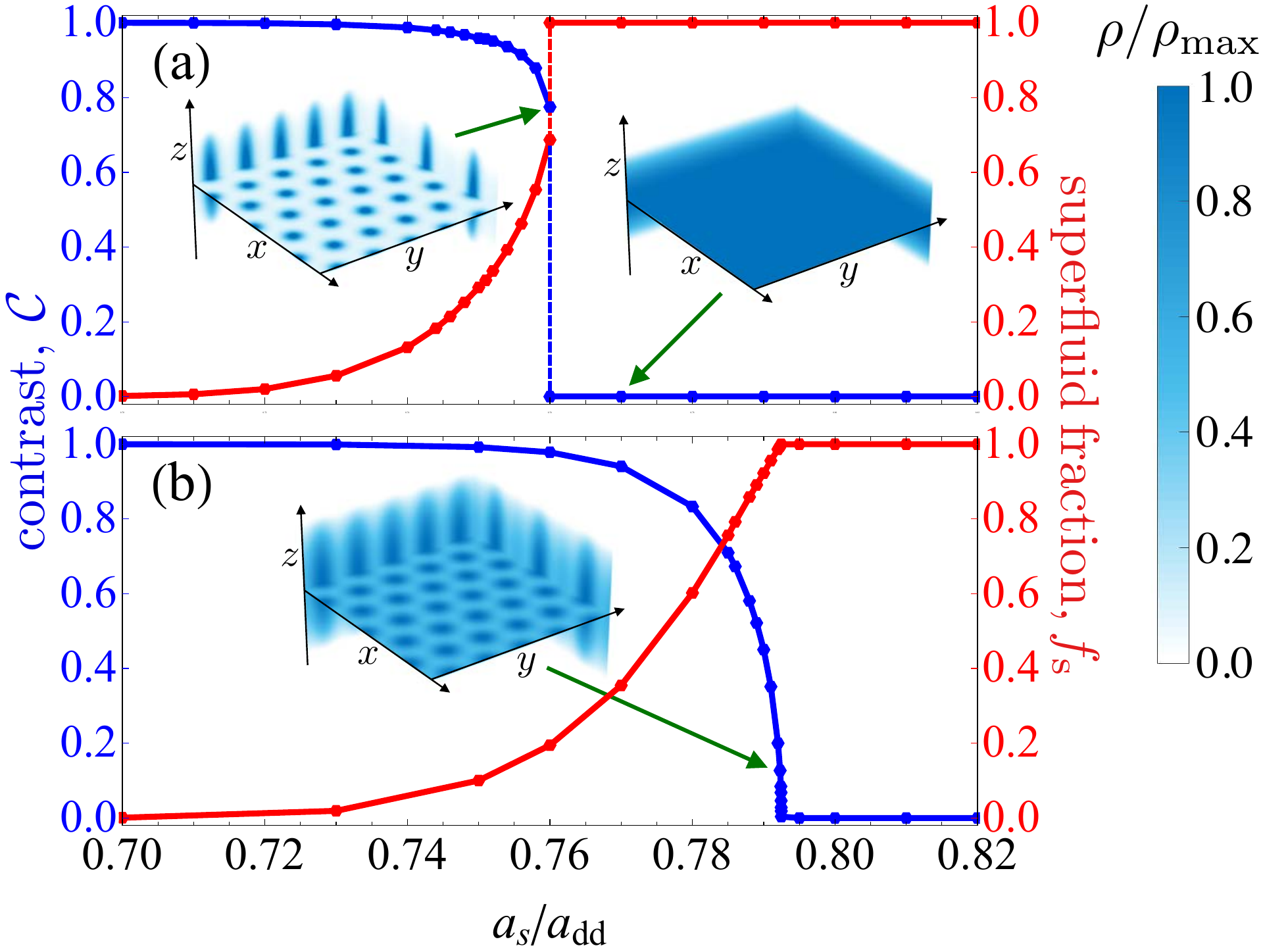}
\caption{(color online) Density contrast and superfluid fraction as function of $a_{\rm s}/a_{\rm dd}$ for a trap frequency $\omega_z=0.08$ and two different densities $\rho_{\rm 2D}=62.5$ (a) and $\rho_{\rm 2D}=156$ (b). The insets show the corresponding density profile at the respective parameters indicated by the green arrows, where the color scale indicates the density $\rho({\bf r})$ relative to the maximum density $\rho_{\rm max}=\max_{\bf r}\rho({\bf r})$. In (b), the contrast undergoes a continuous transition, giving rise to weakly modulated supersolid states with a high superfluidity.
}
\label{fig2}
\end{figure}

As a suitable parameter to detect spatial ordering across the fluid-solid phase transition we consider the density contrast $\mathcal{C}=(|\psi_{\rm max}|^2-|\psi_{\rm min}|^2)/(|\psi_{\rm max}|^2+|\psi_{\rm min}|^2)$, defined through the maximum and minimum density at the trap center, $z=0$.
$\mathcal{C}$ vanishes in the superfluid phase and approaches unity for a crystal of disconnected droplets with vanishing superfluidity. In the present case of weak interactions, the superfluidity can be accurately estimated by Leggett's upper bound on the superfluid fraction~\cite{Leggett:PRL:1970}
\begin{equation}\label{eq:SF}
 f_{\rm s} = \min_\theta\left[\int\frac{(\int {\rm d} x)^2}{\int|\psi(\bar x,\bar y,z)|^{-2}{\rm d} x}{\rm d} y{\rm d}z\right].
 \end{equation}
\refeq{eq:SF} generalizes Leggett's derivation~\cite{Leggett:PRL:1970} for the superfluid response of a rotating system to a linear Galilean boost, whereby we take the minimum with respect to all possible directions defined by the angle $\theta$ with $\bar x=x\cos\theta-y\sin\theta$ and $\bar y=x\sin\theta+y\cos\theta$.

\reffig{fig2}(a) shows these two quantities as we vary the relative strength $a_{\rm s}/a_{\rm dd}$ of the zero-range and dipolar interactions for a fixed trap frequency and a fixed average two-dimensional density $\rho_{\rm 2D}=\int|\psi({\bf r})|^2{\rm d}^3r/(\int{\rm d}x{\rm d}y)$. For large values of $a_{\rm s}/a_{\rm dd}$, zero-range repulsion dominates such that the lowest energy state is an infinitely extended superfluid with $\mathcal{C}=0$ and $f_{\rm s}=1$. However, as we decrease the ratio $a_{\rm s}/a_{\rm dd}$  dipolar interactions cause an abrupt transition to a droplet crystal with a high contrast. This first-order phase transition is consistent with the scenario described by Gross \cite{Gross:PR:1957,Gross:AnnPhys:1958} for the potential formation of a supersolid within a mean field picture based on the Gross-Pitaevskii equation for interacting Bosons and is typical for quantum as well as classical crystallization phenomena in more than one spatial dimension \cite{pattern_RMP}.

Surprisingly, this behaviour changes profoundly for a different choice of the atomic density and is replaced by a second-order quantum phase transition. As shown in Fig.\ref{fig2}(b), the abrupt transition to a droplet crystal ground state is replaced with the gradual emergence of a triangular density wave at the transition point, around which one can find supersolid states with a broken continuous translational symmetry and near unit superfluidity. 

To gain a better intuition for the origin of this behaviour, we consider parameter values around the second-order phase transition and approximate the ground state by a two-dimensional density wave described by
\begin{equation}
    \rho({\bf r}_{\perp},z)=\rho_0(z)\left( 1 + A \sum_{j=1}^3  \cos ({\bm k}_{j}{\bf r}) \right).
    \label{eq:pert_TF}
\end{equation} 
Here, $A$ denotes the small amplitude of the triangular density modulation defined by the three wave vectors, ${\bm k}_{j}$, that form an equilateral triangle in the transverse plane with ${\bm k}_{1}+{\bm k}_{2}+{\bm k}_{3}=0$ and $|{\bm k}_{j}|=k$. Our numerical solutions are well approximated by this ansatz combined with a Thomas-Fermi profile
\begin{equation}\label{eq:TF}
    \rho_0(z) = \frac{3\rho_{\rm 2D}}{4\sigma_z}\left(1-\frac{z^2}{\sigma^2_z}\right), 
    {\ }
    \sigma_z = \left(\frac{\rho_{\rm  2D}(a_{\rm s}/a_{\rm dd}+2)}{2\omega_z^2}\right)^{1/3}
\end{equation}
for the axial density. Upon substituting these expressions into Eqs.~(\ref{eq:energy}) and (\ref{eq:Epot}) we obtain 
\begin{equation}\label{eq:lin_pert}
\frac{\Delta E}{N}=a^{(2)}A^2+a^{(3)}A^3+a^{(4)}A^4
\end{equation}
for the energy difference per atom between the density wave state described by~\refeq{eq:pert_TF} and the unmodulated superfluid with $A=0$. The nature of the transition between them is determined by the cubic term 
\begin{equation}\label{eq:a3}
a^{(3)}=-\frac{3}{32}\left[k^2 -\frac{15\pi \gamma}{32}\left( \frac{3\rho_{\rm 2D}}{4\sigma_z}\right)^{3/2} \right].
\end{equation}
The condensate ground state is readily found by minimizing Eq.~(\ref{eq:lin_pert}) with respect to $A$ and $k$. The resulting ground state behavior is critically determined by the magnitude and sign of $a^{(3)}$. Without the LHY correction, $\gamma=0$, $a^{(3)}$ is always negative such that $A\geq0$ in the condensate ground state. More importantly, the presence of a finite cubic term $a^{(3)}< 0$ implies that any transition to a symmetry broken state must be a first-order transition \cite{Landau:1937}, as illustrated in Fig.\ref{fig1}(d). This scenario is prototypical for crystallization phenomena in more than one spatial dimension for both classical and quantum systems, such as supersolid formation in condensates of soft-core Bosons \cite{Tommaso:PRA:2013,Cinti_2014} or Bose Einstein condensates in optical cavities \cite{cavityBEC}. 

\begin{figure}[t!]
\centering
  \includegraphics[width=\columnwidth]{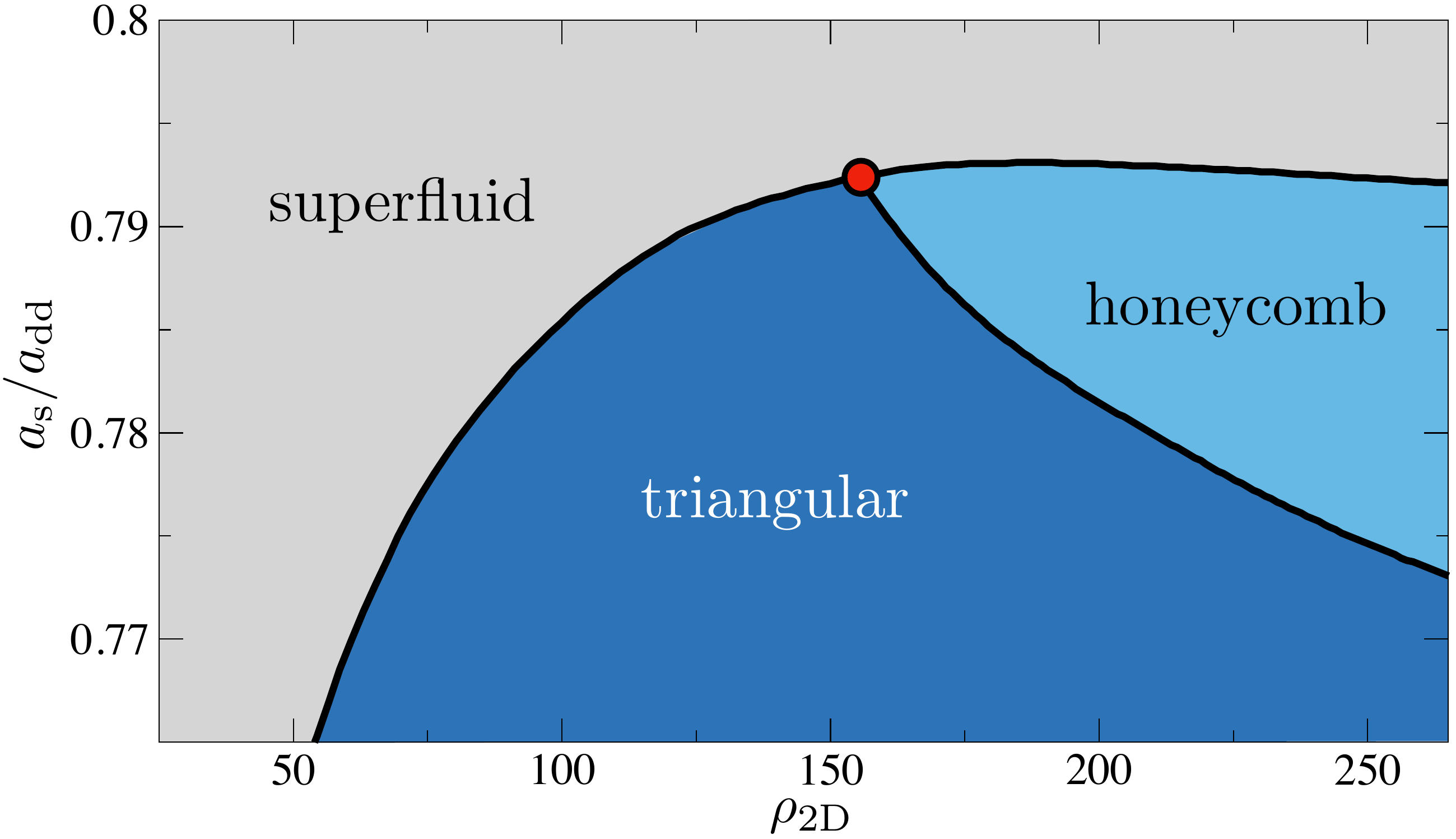}
\caption{(color online) Ground state phase diagram for a fixed trap frequency $\omega_z=0.08$. The lines mark the numerically determined first-order phase boundaries, while the red dot indicates the critical where coexistence of all three ground states terminates.
}
\label{fig3}
\end{figure}

On the contrary, in the present case, beyond meanfield effects can profoundly change this picture, as the LHY correction competes with the kinetic energy contribution to the cubic term in Eq.~(\ref{eq:lin_pert}). In fact, $a^{(3)}$ vanishes if the spatial modulation frequency of the  symmetry-broken ground state satisfies 
\begin{equation}
    k^2 = \frac{15\pi\gamma}{32}\left(\frac{3\rho_{\rm 2D}}{4\sigma_z}\right)^{3/2}.
    \label{eq:period}
\end{equation}
In this case the total energy becomes an even function of the order parameter $A$ and the associated phase transition turns into a second-order transition, consistent with the numerical results shown in \reffig{fig2}(b). The different scenarios are illustrated in \reffig{fig1} where we plot the energy difference as function of $A$ across the phase transition schematically. Prior to the fluid-solid transition, the presence of the cubic term gives rise to a local energy minimum that eventually becomes negative at the transition to a symmetry-broken ground state with a finite modulation amplitude. However, under the condition (\ref{eq:period}), no such local minimum can exist prior to crystallization and the symmetry-broken state emerges with a vanishing modulation amplitude that grows gradually upon crossing the second-order phase transition.

This simple variational approach provides a consistent understanding of the underlying ground state phase diagram shown in Fig.\ref{fig3}. Equation (\ref{eq:lin_pert}) and the above discussion suggest that the effects of the LHY correction on the nature of the phase transition are well controlled  by the atomic density. Indeed, upon varying $\rho_{\rm 2D}$, one finds a second-order quantum phase transition at a density $\rho^{({\rm cr})}_{\rm 2D}$, as indicated by the red dot in Fig.\ref{fig3}. It corresponds to a density of $\rho^{({\rm cr})}_{\rm 2D}=156$ in good agreement with the approximate variational result of $\rho^{({\rm cr})}_{\rm 2D}=140$. On the other hand, the system is significantly less sensitive to the strength of the longitudinal trap. In fact, doubling the trap frequency used in \reffig{fig3} merely increases the critical density by less than $10\%$. Consequently, supersolid ground states with a high superfluidity can be generated for a broad range of trap frequencies upon properly choosing the dimensionless atomic density $\rho_{\rm 2D}\approx 150$.

\begin{figure}[t!]
\centering
  \includegraphics[width=0.99\columnwidth]{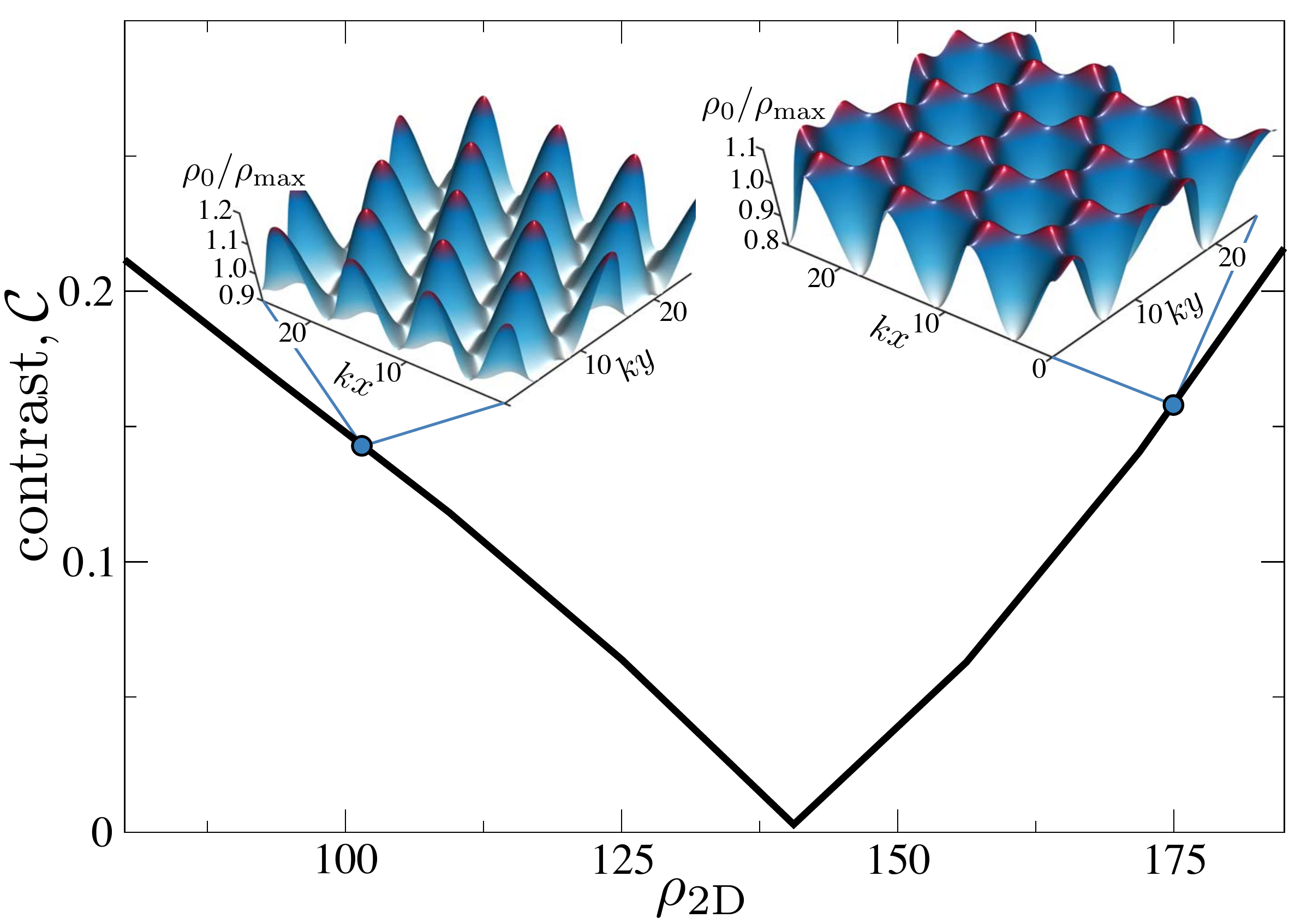}
\caption{(color online) Contrast as a function of the condensate density $\rho_{\rm 2D}$ along the melting line as obtained from eq.(\ref{eq:TF}) and (\ref{eq:lin_pert}) for $\omega_z=0.08$. The insets illustrate the density profile [eq.(\ref{eq:pert_TF})] of the triangular and honeycomb supersolid states on both sides of the critical point where $\mathcal{C}=0$. The depicted surfaces show the density $\rho_0(x,y)=\rho({\bf r})|_{z=0}$ at the trap center.}
\label{fig4}
\end{figure}

The LHY-term in Eq.(\ref{eq:a3}) also makes a sign change of $a^{(3)}$ possible, which yields a new ground state with $A<0$.
Interestingly, this configuration does not lead into the formation of a regular droplet crystal, but instead establishes a honeycomb structure [see Fig.\ref{fig1}(a)] that behaves differently from such previously discussed \cite{Henkel:PRL:2010,CA2,Blakie:PRL:2018} ordered states. Along the melting line (see Fig.\ref{fig3}), the two symmetry broken phases are separated by the critical density $\rho_{\rm 2D}^{({\rm cr})}$ and feature an increasing contrast $\mathcal{C}$ away from it, as shown in Fig.\ref{fig4}. However, the honeycomb ground state does not disintegrate into a crystal of separate droplets but maintains a connected density pattern. This corresponds to a remarkable supersolid ground state that features a high contrast $\mathcal{C}\approx1$, while maintaining a near-perfect superfluid flow along the boundaries of the honeycomb structure.
The droplet crystal phase and the superfluid honeycomb are separated by a first-order transition line that ends at the critical point with $\rho_{\rm 2D}^{({\rm cr})}$ where three first-order transition lines meet and the coexistence of all three ground state phases terminates. 

The parameters considered above fall well into the regime of achievable conditions in current experiments~\cite{Pfau:nature:2016,Pfau:nature2:2016,roton1,Tanzi,1D_2}.
For example dysprosium atoms have dipolar interactions with $a_{\rm dd}\approx7$nm and offer a broadly tunable s-wave scattering length, covering our considered range of $a_{\rm dd}/a_{\rm s}$. With an atomic mass of $m=2.7\times10^{-25}$kg, the defined spatial and temporal unit of dysprosium condensates is $\ell=0.26\mu$m and $m\ell^2/\hbar=0.18$ms, respectively. The considered dimensionless trapping frequencies of $\omega_z\approx0.1$ thus corresponds to a value of $560$Hz that is  well achievable in cold atom experiments. Around the critical point of Fig.\ref{fig3} the dimensionless values of  $\rho_{\rm 2D}\approx150$ and $a_{\rm s}/a_{\rm dd}\approx0.78$ imply a  condensate thickness of $\sigma_z\approx7\mu$m and an associated peak densities of $3\rho_{\rm 2D}/4\sigma_z\approx2\times 10^{14}$cm$^{-3}$ that are typical for cold atom experiments. Finally, \refeq{eq:period} yields an estimated periodicity of $2\pi/k\approx3.7\mu$m that is sufficiently small to accommodate extended supersolid states with reasonable atom numbers and can be further decreased by tightening the longitudinal confinement.

In conclusion, we have investigated the two-dimensional crystallization of partially confined dipolar Bose Einstein condensates. It turns out that the significant role of quantum fluctuations in such systems can profoundly affect the underlying crystallization mechanism as compared to current scenarios for finite-range interacting Bosons. Specifically, we have shown that quantum fluctuations lead to the emergence of a critical point around which spatial symmetry breaking takes place via the gradual growths of a density wave and thereby facilitates the formation of supersolid ground states with a high superfluidity. Moreover, we have demonstrated a new supersolid ground state that is solely stabilized by quantum fluctuations and features a maximally modulated density while maintaining near perfect superfluidity. 

The quantum phase diagram, described in this work, may also suggest viable schemes for the experimental preparation of symmetry-broken low-energy states via slow parameter changes in the vicinity of the continuous phase transition. In addition to exploring such dynamical aspects, the present results raise a number of further outstanding questions for future work. These include the thermodynamics of the described crystallization scenario to elucidate the effects and interplay of quantum and thermal fluctuation as well as the fate of the different supersolid phases upon approaching the strong-interaction regime where quantum fluctuations and lattice defects may play a more prominent role. 

We thank Georg Bruun for valuable discussions and acknowledge funding by the DNRF through a Niels Bohr Professorship.


\begin{thebibliography}{42}%
\makeatletter
\providecommand \@ifxundefined [1]{%
 \@ifx{#1\undefined}
}%
\providecommand \@ifnum [1]{%
 \ifnum #1\expandafter \@firstoftwo
 \else \expandafter \@secondoftwo
 \fi
}%
\providecommand \@ifx [1]{%
 \ifx #1\expandafter \@firstoftwo
 \else \expandafter \@secondoftwo
 \fi
}%
\providecommand \natexlab [1]{#1}%
\providecommand \enquote  [1]{``#1''}%
\providecommand \bibnamefont  [1]{#1}%
\providecommand \bibfnamefont [1]{#1}%
\providecommand \citenamefont [1]{#1}%
\providecommand \href@noop [0]{\@secondoftwo}%
\providecommand \href [0]{\begingroup \@sanitize@url \@href}%
\providecommand \@href[1]{\@@startlink{#1}\@@href}%
\providecommand \@@href[1]{\endgroup#1\@@endlink}%
\providecommand \@sanitize@url [0]{\catcode `\\12\catcode `\$12\catcode
  `\&12\catcode `\#12\catcode `\^12\catcode `\_12\catcode `\%12\relax}%
\providecommand \@@startlink[1]{}%
\providecommand \@@endlink[0]{}%
\providecommand \url  [0]{\begingroup\@sanitize@url \@url }%
\providecommand \@url [1]{\endgroup\@href {#1}{\urlprefix }}%
\providecommand \urlprefix  [0]{URL }%
\providecommand \Eprint [0]{\href }%
\providecommand \doibase [0]{http://dx.doi.org/}%
\providecommand \selectlanguage [0]{\@gobble}%
\providecommand \bibinfo  [0]{\@secondoftwo}%
\providecommand \bibfield  [0]{\@secondoftwo}%
\providecommand \translation [1]{[#1]}%
\providecommand \BibitemOpen [0]{}%
\providecommand \bibitemStop [0]{}%
\providecommand \bibitemNoStop [0]{.\EOS\space}%
\providecommand \EOS [0]{\spacefactor3000\relax}%
\providecommand \BibitemShut  [1]{\csname bibitem#1\endcsname}%
\let\auto@bib@innerbib\@empty
\bibitem [{\citenamefont {Gross}(1957)}]{Gross:PR:1957}%
  \BibitemOpen
  \bibfield  {author} {\bibinfo {author} {\bibfnamefont {E.~P.}\ \bibnamefont
  {Gross}},\ }\href@noop {} {\bibfield  {journal} {\bibinfo  {journal} {Phys.
  Rev.}\ }\textbf {\bibinfo {volume} {106}},\ \bibinfo {pages} {161} (\bibinfo
  {year} {1957})}\BibitemShut {NoStop}%
\bibitem [{\citenamefont {Andreev}\ and\ \citenamefont
  {Lifshitz}(1969)}]{Andreev:JETP:1969}%
  \BibitemOpen
  \bibfield  {author} {\bibinfo {author} {\bibfnamefont {A.~F.}\ \bibnamefont
  {Andreev}}\ and\ \bibinfo {author} {\bibfnamefont {I.~M.}\ \bibnamefont
  {Lifshitz}},\ }\href@noop {} {\bibfield  {journal} {\bibinfo  {journal} {Sov.
  Phys. JETP}\ }\textbf {\bibinfo {volume} {29}},\ \bibinfo {pages} {1107}
  (\bibinfo {year} {1969})}\BibitemShut {NoStop}%
\bibitem [{\citenamefont {Chester}(1970)}]{Chester:PRA:1970}%
  \BibitemOpen
  \bibfield  {author} {\bibinfo {author} {\bibfnamefont {G.~V.}\ \bibnamefont
  {Chester}},\ }\href@noop {} {\bibfield  {journal} {\bibinfo  {journal} {Phys.
  Rev. A}\ }\textbf {\bibinfo {volume} {2}},\ \bibinfo {pages} {256} (\bibinfo
  {year} {1970})}\BibitemShut {NoStop}%
\bibitem [{\citenamefont {Leggett}(1970)}]{Leggett:PRL:1970}%
  \BibitemOpen
  \bibfield  {author} {\bibinfo {author} {\bibfnamefont {A.~J.}\ \bibnamefont
  {Leggett}},\ }\href@noop {} {\bibfield  {journal} {\bibinfo  {journal} {Phys.
  Rev. Lett.}\ }\textbf {\bibinfo {volume} {25}},\ \bibinfo {pages} {1543}
  (\bibinfo {year} {1970})}\BibitemShut {NoStop}%
\bibitem [{\citenamefont {Meisel}(1992)}]{suso_Rev}%
  \BibitemOpen
  \bibfield  {author} {\bibinfo {author} {\bibfnamefont {M.~W.}\ \bibnamefont
  {Meisel}},\ }\href {\doibase https://doi.org/10.1016/0921-4526(92)90186-V}
  {\bibfield  {journal} {\bibinfo  {journal} {Physica B: Condensed Matter}\
  }\textbf {\bibinfo {volume} {178}},\ \bibinfo {pages} {121 } (\bibinfo {year}
  {1992})}%
\bibitem [{\citenamefont {Boninsegni}\ and\ \citenamefont
  {Prokof'ev}(2012)}]{suso_RMP}%
  \BibitemOpen
  \bibfield  {author} {\bibinfo {author} {\bibfnamefont {M.}~\bibnamefont
  {Boninsegni}}\ and\ \bibinfo {author} {\bibfnamefont {N.~V.}\ \bibnamefont
  {Prokof'ev}},\ }\href {\doibase 10.1103/RevModPhys.84.759} {\bibfield
  {journal} {\bibinfo  {journal} {Rev. Mod. Phys.}\ }\textbf {\bibinfo {volume}
  {84}},\ \bibinfo {pages} {759} (\bibinfo {year} {2012})}\BibitemShut
  {NoStop}%
\bibitem [{\citenamefont {Gopalakrishnan}\ \emph {et~al.}(2009)\citenamefont
  {Gopalakrishnan}, \citenamefont {Lev},\ and\ \citenamefont {Goldbart}}]{CA1}%
  \BibitemOpen
  \bibfield  {author} {\bibinfo {author} {\bibfnamefont {S.}~\bibnamefont
  {Gopalakrishnan}}, \bibinfo {author} {\bibfnamefont {B.~L.}\ \bibnamefont
  {Lev}}, \ and\ \bibinfo {author} {\bibfnamefont {P.~M.}\ \bibnamefont
  {Goldbart}},\ }\href {https://doi.org/10.1038/nphys1403} {\bibfield
  {journal} {\bibinfo  {journal} {Nature Physics}\ }\textbf {\bibinfo {volume}
  {5}},\ \bibinfo {pages} {845} (\bibinfo {year} {2009})}\BibitemShut
  {NoStop}%
\bibitem [{\citenamefont {Henkel}\ \emph {et~al.}(2010)\citenamefont {Henkel},
  \citenamefont {Nath},\ and\ \citenamefont {Pohl}}]{Henkel:PRL:2010}%
  \BibitemOpen
  \bibfield  {author} {\bibinfo {author} {\bibfnamefont {N.}~\bibnamefont
  {Henkel}}, \bibinfo {author} {\bibfnamefont {R.}~\bibnamefont {Nath}}, \ and\
  \bibinfo {author} {\bibfnamefont {T.}~\bibnamefont {Pohl}},\ }\href@noop {}
  {\bibfield  {journal} {\bibinfo  {journal} {Phys. Rev. Lett.}\ }\textbf
  {\bibinfo {volume} {104}},\ \bibinfo {pages} {195302} (\bibinfo {year}
  {2010})}\BibitemShut {NoStop}%
\bibitem [{\citenamefont {Cinti}\ \emph {et~al.}(2010)\citenamefont {Cinti},
  \citenamefont {Jain}, \citenamefont {Boninsegni}, \citenamefont {Micheli},
  \citenamefont {Zoller},\ and\ \citenamefont {Pupillo}}]{CA2}%
  \BibitemOpen
  \bibfield  {author} {\bibinfo {author} {\bibfnamefont {F.}~\bibnamefont
  {Cinti}}, \bibinfo {author} {\bibfnamefont {P.}~\bibnamefont {Jain}},
  \bibinfo {author} {\bibfnamefont {M.}~\bibnamefont {Boninsegni}}, \bibinfo
  {author} {\bibfnamefont {A.}~\bibnamefont {Micheli}}, \bibinfo {author}
  {\bibfnamefont {P.}~\bibnamefont {Zoller}}, \ and\ \bibinfo {author}
  {\bibfnamefont {G.}~\bibnamefont {Pupillo}},\ }\href {\doibase
  10.1103/PhysRevLett.105.135301} {\bibfield  {journal} {\bibinfo  {journal}
  {Phys. Rev. Lett.}\ }\textbf {\bibinfo {volume} {105}},\ \bibinfo {pages}
  {135301} (\bibinfo {year} {2010})}\BibitemShut {NoStop}%
\bibitem [{\citenamefont {Cinti}\ \emph
  {et~al.}(2014{\natexlab{a}})\citenamefont {Cinti}, \citenamefont
  {Macr{\`\i}}, \citenamefont {Lechner}, \citenamefont {Pupillo},\ and\
  \citenamefont {Pohl}}]{CA3}%
  \BibitemOpen
  \bibfield  {author} {\bibinfo {author} {\bibfnamefont {F.}~\bibnamefont
  {Cinti}}, \bibinfo {author} {\bibfnamefont {T.}~\bibnamefont {Macr{\`\i}}},
  \bibinfo {author} {\bibfnamefont {W.}~\bibnamefont {Lechner}}, \bibinfo
  {author} {\bibfnamefont {G.}~\bibnamefont {Pupillo}}, \ and\ \bibinfo
  {author} {\bibfnamefont {T.}~\bibnamefont {Pohl}},\ }\href
  {https://doi.org/10.1038/ncomms4235} {\bibfield  {journal} {\bibinfo
  {journal} {Nature Communications}\ }\textbf {\bibinfo {volume} {5}},\
  \bibinfo {pages} {3235} (\bibinfo {year}
  {2014}{\natexlab{a}})}\BibitemShut {NoStop}%
\bibitem [{\citenamefont {Li}\ \emph {et~al.}(2013)\citenamefont {Li},
  \citenamefont {Martone}, \citenamefont {Pitaevskii},\ and\ \citenamefont
  {Stringari}}]{CA4}%
  \BibitemOpen
  \bibfield  {author} {\bibinfo {author} {\bibfnamefont {Y.}~\bibnamefont
  {Li}}, \bibinfo {author} {\bibfnamefont {G.~I.}\ \bibnamefont {Martone}},
  \bibinfo {author} {\bibfnamefont {L.~P.}\ \bibnamefont {Pitaevskii}}, \ and\
  \bibinfo {author} {\bibfnamefont {S.}~\bibnamefont {Stringari}},\ }\href
  {\doibase 10.1103/PhysRevLett.110.235302} {\bibfield  {journal} {\bibinfo
  {journal} {Phys. Rev. Lett.}\ }\textbf {\bibinfo {volume} {110}},\ \bibinfo
  {pages} {235302} (\bibinfo {year} {2013})}\BibitemShut {NoStop}%
\bibitem [{\citenamefont {L{\'e}onard}\ \emph {et~al.}(2017)\citenamefont
  {L{\'e}onard}, \citenamefont {Morales}, \citenamefont {Zupancic},
  \citenamefont {Esslinger},\ and\ \citenamefont
  {Donner}}]{Donner:nature:2017}%
  \BibitemOpen
  \bibfield  {author} {\bibinfo {author} {\bibfnamefont {J.}~\bibnamefont
  {L{\'e}onard}}, \bibinfo {author} {\bibfnamefont {A.}~\bibnamefont
  {Morales}}, \bibinfo {author} {\bibfnamefont {P.}~\bibnamefont {Zupancic}},
  \bibinfo {author} {\bibfnamefont {T.}~\bibnamefont {Esslinger}}, \ and\
  \bibinfo {author} {\bibfnamefont {T.}~\bibnamefont {Donner}},\ }\href@noop {}
  {\bibfield  {journal} {\bibinfo  {journal} {Nature}\ }\textbf {\bibinfo
  {volume} {543}},\ \bibinfo {pages} {87} (\bibinfo {year} {2017})}\BibitemShut
  {NoStop}%
\bibitem [{\citenamefont {Li}\ \emph {et~al.}(2017)\citenamefont {Li},
  \citenamefont {Lee}, \citenamefont {Huang}, \citenamefont {Burchesky},
  \citenamefont {Shteynas}, \citenamefont {Top}, \citenamefont {Jamison},\ and\
  \citenamefont {Ketterle}}]{Ketterle:nature:2017}%
  \BibitemOpen
  \bibfield  {author} {\bibinfo {author} {\bibfnamefont {J.}~\bibnamefont
  {Li}}, \bibinfo {author} {\bibfnamefont {J.}~\bibnamefont {Lee}}, \bibinfo
  {author} {\bibfnamefont {W.}~\bibnamefont {Huang}}, \bibinfo {author}
  {\bibfnamefont {S.}~\bibnamefont {Burchesky}}, \bibinfo {author}
  {\bibfnamefont {B.}~\bibnamefont {Shteynas}}, \bibinfo {author}
  {\bibfnamefont {F.~C.}\ \bibnamefont {Top}}, \bibinfo {author} {\bibfnamefont
  {A.~O.}\ \bibnamefont {Jamison}}, \ and\ \bibinfo {author} {\bibfnamefont
  {W.}~\bibnamefont {Ketterle}},\ }\href@noop {} {\bibfield  {journal}
  {\bibinfo  {journal} {Nature}\ }\textbf {\bibinfo {volume} {543}},\ \bibinfo
  {pages} {91} (\bibinfo {year} {2017})}\BibitemShut {NoStop}%
\bibitem [{\citenamefont {Lu}\ \emph {et~al.}(2011)\citenamefont {Lu},
  \citenamefont {Burdick}, \citenamefont {Youn},\ and\ \citenamefont
  {Lev}}]{Dy1}%
  \BibitemOpen
  \bibfield  {author} {\bibinfo {author} {\bibfnamefont {M.}~\bibnamefont
  {Lu}}, \bibinfo {author} {\bibfnamefont {N.~Q.}\ \bibnamefont {Burdick}},
  \bibinfo {author} {\bibfnamefont {S.~H.}\ \bibnamefont {Youn}}, \ and\
  \bibinfo {author} {\bibfnamefont {B.~L.}\ \bibnamefont {Lev}},\ }\href
  {\doibase 10.1103/PhysRevLett.107.190401} {\bibfield  {journal} {\bibinfo
  {journal} {Phys. Rev. Lett.}\ }\textbf {\bibinfo {volume} {107}},\ \bibinfo
  {pages} {190401} (\bibinfo {year} {2011})}\BibitemShut {NoStop}%
\bibitem [{\citenamefont {Aikawa}\ \emph {et~al.}(2012)\citenamefont {Aikawa},
  \citenamefont {Frisch}, \citenamefont {Mark}, \citenamefont {Baier},
  \citenamefont {Rietzler}, \citenamefont {Grimm},\ and\ \citenamefont
  {Ferlaino}}]{Er1}%
  \BibitemOpen
  \bibfield  {author} {\bibinfo {author} {\bibfnamefont {K.}~\bibnamefont
  {Aikawa}}, \bibinfo {author} {\bibfnamefont {A.}~\bibnamefont {Frisch}},
  \bibinfo {author} {\bibfnamefont {M.}~\bibnamefont {Mark}}, \bibinfo {author}
  {\bibfnamefont {S.}~\bibnamefont {Baier}}, \bibinfo {author} {\bibfnamefont
  {A.}~\bibnamefont {Rietzler}}, \bibinfo {author} {\bibfnamefont
  {R.}~\bibnamefont {Grimm}}, \ and\ \bibinfo {author} {\bibfnamefont
  {F.}~\bibnamefont {Ferlaino}},\ }\href {\doibase
  10.1103/PhysRevLett.108.210401} {\bibfield  {journal} {\bibinfo  {journal}
  {Phys. Rev. Lett.}\ }\textbf {\bibinfo {volume} {108}},\ \bibinfo {pages}
  {210401} (\bibinfo {year} {2012})}\BibitemShut {NoStop}%
\bibitem [{\citenamefont {Chomaz}\ \emph {et~al.}(2018)\citenamefont {Chomaz},
  \citenamefont {van Bijnen}, \citenamefont {Petter}, \citenamefont {Faraoni},
  \citenamefont {Baier}, \citenamefont {Becher}, \citenamefont {Mark},
  \citenamefont {W{\"a}chtler}, \citenamefont {Santos},\ and\ \citenamefont
  {Ferlaino}}]{roton1}%
  \BibitemOpen
  \bibfield  {author} {\bibinfo {author} {\bibfnamefont {L.}~\bibnamefont
  {Chomaz}}, \bibinfo {author} {\bibfnamefont {R.~M.~W.}\ \bibnamefont {van
  Bijnen}}, \bibinfo {author} {\bibfnamefont {D.}~\bibnamefont {Petter}},
  \bibinfo {author} {\bibfnamefont {G.}~\bibnamefont {Faraoni}}, \bibinfo
  {author} {\bibfnamefont {S.}~\bibnamefont {Baier}}, \bibinfo {author}
  {\bibfnamefont {J.~H.}\ \bibnamefont {Becher}}, \bibinfo {author}
  {\bibfnamefont {M.~J.}\ \bibnamefont {Mark}}, \bibinfo {author}
  {\bibfnamefont {F.}~\bibnamefont {W{\"a}chtler}}, \bibinfo {author}
  {\bibfnamefont {L.}~\bibnamefont {Santos}}, \ and\ \bibinfo {author}
  {\bibfnamefont {F.}~\bibnamefont {Ferlaino}},\ }\href {\doibase
  10.1038/s41567-018-0054-7} {\bibfield  {journal} {\bibinfo  {journal} {Nature
  Physics}\ }\textbf {\bibinfo {volume} {14}},\ \bibinfo {pages} {442}
  (\bibinfo {year} {2018})}\BibitemShut {NoStop}%
\bibitem [{\citenamefont {Landau}(1949)}]{Landau:PR:1949}%
  \BibitemOpen
  \bibfield  {author} {\bibinfo {author} {\bibfnamefont {L.}~\bibnamefont
  {Landau}},\ }\href@noop {} {\bibfield  {journal} {\bibinfo  {journal} {Phys.
  Rev.}\ }\textbf {\bibinfo {volume} {75}},\ \bibinfo {pages} {884} (\bibinfo
  {year} {1949})}\BibitemShut {NoStop}%
\bibitem [{\citenamefont {Kadau}\ \emph {et~al.}(2016)\citenamefont {Kadau},
  \citenamefont {Schmitt}, \citenamefont {Wenzel}, \citenamefont {Wink},
  \citenamefont {Maier}, \citenamefont {Ferrier-Barbut},\ and\ \citenamefont
  {Pfau}}]{Pfau:nature:2016}%
  \BibitemOpen
  \bibfield  {author} {\bibinfo {author} {\bibfnamefont {H.}~\bibnamefont
  {Kadau}}, \bibinfo {author} {\bibfnamefont {M.}~\bibnamefont {Schmitt}},
  \bibinfo {author} {\bibfnamefont {M.}~\bibnamefont {Wenzel}}, \bibinfo
  {author} {\bibfnamefont {C.}~\bibnamefont {Wink}}, \bibinfo {author}
  {\bibfnamefont {T.}~\bibnamefont {Maier}}, \bibinfo {author} {\bibfnamefont
  {I.}~\bibnamefont {Ferrier-Barbut}}, \ and\ \bibinfo {author} {\bibfnamefont
  {T.}~\bibnamefont {Pfau}},\ }\href@noop {} {\bibfield  {journal} {\bibinfo
  {journal} {Nature}\ }\textbf {\bibinfo {volume} {530}},\ \bibinfo {pages}
  {194} (\bibinfo {year} {2016})}\BibitemShut {NoStop}%
\bibitem [{\citenamefont {Schmitt}\ \emph {et~al.}(2016)\citenamefont
  {Schmitt}, \citenamefont {Wenzel}, \citenamefont {B{\"o}ttcher},
  \citenamefont {Ferrier-Barbut},\ and\ \citenamefont
  {Pfau}}]{Pfau:nature2:2016}%
  \BibitemOpen
  \bibfield  {author} {\bibinfo {author} {\bibfnamefont {M.}~\bibnamefont
  {Schmitt}}, \bibinfo {author} {\bibfnamefont {M.}~\bibnamefont {Wenzel}},
  \bibinfo {author} {\bibfnamefont {F.}~\bibnamefont {B{\"o}ttcher}}, \bibinfo
  {author} {\bibfnamefont {I.}~\bibnamefont {Ferrier-Barbut}}, \ and\ \bibinfo
  {author} {\bibfnamefont {T.}~\bibnamefont {Pfau}},\ }\href@noop {} {\bibfield
   {journal} {\bibinfo  {journal} {Nature}\ }\textbf {\bibinfo {volume}
  {539}},\ \bibinfo {pages} {259} (\bibinfo {year} {2016})}\BibitemShut
  {NoStop}%
\bibitem [{\citenamefont {Chomaz}\ \emph {et~al.}(2016)\citenamefont {Chomaz},
  \citenamefont {Baier}, \citenamefont {Petter}, \citenamefont {Mark},
  \citenamefont {W\"achtler}, \citenamefont {Santos},\ and\ \citenamefont
  {Ferlaino}}]{Santos:PRX:2016}%
  \BibitemOpen
  \bibfield  {author} {\bibinfo {author} {\bibfnamefont {L.}~\bibnamefont
  {Chomaz}}, \bibinfo {author} {\bibfnamefont {S.}~\bibnamefont {Baier}},
  \bibinfo {author} {\bibfnamefont {D.}~\bibnamefont {Petter}}, \bibinfo
  {author} {\bibfnamefont {M.~J.}\ \bibnamefont {Mark}}, \bibinfo {author}
  {\bibfnamefont {F.}~\bibnamefont {W\"achtler}}, \bibinfo {author}
  {\bibfnamefont {L.}~\bibnamefont {Santos}}, \ and\ \bibinfo {author}
  {\bibfnamefont {F.}~\bibnamefont {Ferlaino}},\ }\href@noop {} {\bibfield
  {journal} {\bibinfo  {journal} {Phys. Rev. X}\ }\textbf {\bibinfo {volume}
  {6}},\ \bibinfo {pages} {041039} (\bibinfo {year} {2016})}\BibitemShut
  {NoStop}%
\bibitem [{\citenamefont {Ferrier-Barbut}\ \emph {et~al.}(2016)\citenamefont
  {Ferrier-Barbut}, \citenamefont {Kadau}, \citenamefont {Schmitt},
  \citenamefont {Wenzel},\ and\ \citenamefont
  {Pfau}}]{Ferrier_Barbut:PRL:2016}%
  \BibitemOpen
  \bibfield  {author} {\bibinfo {author} {\bibfnamefont {I.}~\bibnamefont
  {Ferrier-Barbut}}, \bibinfo {author} {\bibfnamefont {H.}~\bibnamefont
  {Kadau}}, \bibinfo {author} {\bibfnamefont {M.}~\bibnamefont {Schmitt}},
  \bibinfo {author} {\bibfnamefont {M.}~\bibnamefont {Wenzel}}, \ and\ \bibinfo
  {author} {\bibfnamefont {T.}~\bibnamefont {Pfau}},\ }\href {\doibase
  10.1103/PhysRevLett.116.215301} {\bibfield  {journal} {\bibinfo  {journal}
  {Phys. Rev. Lett.}\ }\textbf {\bibinfo {volume} {116}},\ \bibinfo {pages}
  {215301} (\bibinfo {year} {2016})}\BibitemShut {NoStop}%
\bibitem [{\citenamefont {W\"achtler}\ and\ \citenamefont
  {Santos}(2016{\natexlab{a}})}]{Santos:PRA:2016}%
  \BibitemOpen
  \bibfield  {author} {\bibinfo {author} {\bibfnamefont {F.}~\bibnamefont
  {W\"achtler}}\ and\ \bibinfo {author} {\bibfnamefont {L.}~\bibnamefont
  {Santos}},\ }\href@noop {} {\bibfield  {journal} {\bibinfo  {journal} {Phys.
  Rev. A}\ }\textbf {\bibinfo {volume} {93}},\ \bibinfo {pages} {061603}
  (\bibinfo {year} {2016}{\natexlab{a}})}\BibitemShut {NoStop}%
\bibitem [{\citenamefont {W\"achtler}\ and\ \citenamefont
  {Santos}(2016{\natexlab{b}})}]{Santos:PRA2:2016}%
  \BibitemOpen
  \bibfield  {author} {\bibinfo {author} {\bibfnamefont {F.}~\bibnamefont
  {W\"achtler}}\ and\ \bibinfo {author} {\bibfnamefont {L.}~\bibnamefont
  {Santos}},\ }\href@noop {} {\bibfield  {journal} {\bibinfo  {journal} {Phys.
  Rev. A}\ }\textbf {\bibinfo {volume} {94}},\ \bibinfo {pages} {043618}
  (\bibinfo {year} {2016}{\natexlab{b}})}\BibitemShut {NoStop}%
\bibitem [{\citenamefont {Bisset}\ \emph {et~al.}(2016)\citenamefont {Bisset},
  \citenamefont {Wilson}, \citenamefont {Baillie},\ and\ \citenamefont
  {Blakie}}]{Bisset:PRA:2016}%
  \BibitemOpen
  \bibfield  {author} {\bibinfo {author} {\bibfnamefont {R.~N.}\ \bibnamefont
  {Bisset}}, \bibinfo {author} {\bibfnamefont {R.~M.}\ \bibnamefont {Wilson}},
  \bibinfo {author} {\bibfnamefont {D.}~\bibnamefont {Baillie}}, \ and\
  \bibinfo {author} {\bibfnamefont {P.~B.}\ \bibnamefont {Blakie}},\
  }\href@noop {} {\bibfield  {journal} {\bibinfo  {journal} {Phys. Rev. A}\
  }\textbf {\bibinfo {volume} {94}},\ \bibinfo {pages} {033619} (\bibinfo
  {year} {2016})}\BibitemShut {NoStop}%
\bibitem [{\citenamefont {Baillie}\ and\ \citenamefont
  {Blakie}(2018)}]{Blakie:PRL:2018}%
  \BibitemOpen
  \bibfield  {author} {\bibinfo {author} {\bibfnamefont {D.}~\bibnamefont
  {Baillie}}\ and\ \bibinfo {author} {\bibfnamefont {P.~B.}\ \bibnamefont
  {Blakie}},\ }\href@noop {} {\bibfield  {journal} {\bibinfo  {journal} {Phys.
  Rev. Lett.}\ }\textbf {\bibinfo {volume} {121}},\ \bibinfo {pages} {195301}
  (\bibinfo {year} {2018})}\BibitemShut {NoStop}%
\bibitem [{\citenamefont {{Roccuzzo}}\ and\ \citenamefont
  {{Ancilotto}}(2018)}]{1D_1}%
  \BibitemOpen
  \bibfield  {author} {\bibinfo {author} {\bibfnamefont {S.~M.}\ \bibnamefont
  {{Roccuzzo}}}\ and\ \bibinfo {author} {\bibfnamefont {F.}~\bibnamefont
  {{Ancilotto}}}},\ \Eprint {http://arxiv.org/abs/1810.12229} {arXiv:1810.12229
  [cond-mat.quant-gas]} \BibitemShut {NoStop}%
\bibitem [{\citenamefont {{Tanzi}}\ \emph {et~al.}(2018)\citenamefont
  {{Tanzi}}, \citenamefont {{Lucioni}}, \citenamefont {{Fam{\`a}}},
  \citenamefont {{Catani}}, \citenamefont {{Fioretti}}, \citenamefont
  {{Gabbanini}}, \citenamefont {{Bisset}}, \citenamefont {{Santos}},\ and\
  \citenamefont {{Modugno}}}]{Tanzi}%
  \BibitemOpen
  \bibfield  {author} {\bibinfo {author} {\bibfnamefont {L.}~\bibnamefont
  {{Tanzi}}}, \bibinfo {author} {\bibfnamefont {E.}~\bibnamefont {{Lucioni}}},
  \bibinfo {author} {\bibfnamefont {F.}~\bibnamefont {{Fam{\`a}}}}, \bibinfo
  {author} {\bibfnamefont {J.}~\bibnamefont {{Catani}}}, \bibinfo {author}
  {\bibfnamefont {A.}~\bibnamefont {{Fioretti}}}, \bibinfo {author}
  {\bibfnamefont {C.}~\bibnamefont {{Gabbanini}}}, \bibinfo {author}
  {\bibfnamefont {R.~N.}\ \bibnamefont {{Bisset}}}, \bibinfo {author}
  {\bibfnamefont {L.}~\bibnamefont {{Santos}}}, \ and\ \bibinfo {author}
  {\bibfnamefont {G.}~\bibnamefont {{Modugno}}},\ }\Eprint
  {http://arxiv.org/abs/1811.02613} {arXiv:1811.02613 [cond-mat.quant-gas]}
  \BibitemShut {NoStop}%
\bibitem [{\citenamefont {{B{\"o}ttcher}}\ \emph {et~al.}(2019)\citenamefont
  {{B{\"o}ttcher}}, \citenamefont {{Schmidt}}, \citenamefont {{Wenzel}},
  \citenamefont {{Hertkorn}}, \citenamefont {{Guo}}, \citenamefont {{Langen}},\
  and\ \citenamefont {{Pfau}}}]{1D_2}%
  \BibitemOpen
  \bibfield  {author} {\bibinfo {author} {\bibfnamefont {F.}~\bibnamefont
  {{B{\"o}ttcher}}}, \bibinfo {author} {\bibfnamefont {J.-N.}\ \bibnamefont
  {{Schmidt}}}, \bibinfo {author} {\bibfnamefont {M.}~\bibnamefont {{Wenzel}}},
  \bibinfo {author} {\bibfnamefont {J.}~\bibnamefont {{Hertkorn}}}, \bibinfo
  {author} {\bibfnamefont {M.}~\bibnamefont {{Guo}}}, \bibinfo {author}
  {\bibfnamefont {T.}~\bibnamefont {{Langen}}}, \ and\ \bibinfo {author}
  {\bibfnamefont {T.}~\bibnamefont {{Pfau}}},\ }\Eprint
  {http://arxiv.org/abs/1901.07982} {arXiv:1901.07982 [cond-mat.quant-gas]}
  \BibitemShut {NoStop}%
\bibitem [{\citenamefont {Cross}\ and\ \citenamefont
  {Hohenberg}(1993)}]{pattern_RMP}%
  \BibitemOpen
  \bibfield  {author} {\bibinfo {author} {\bibfnamefont {M.~C.}\ \bibnamefont
  {Cross}}\ and\ \bibinfo {author} {\bibfnamefont {P.~C.}\ \bibnamefont
  {Hohenberg}},\ }\href {\doibase 10.1103/RevModPhys.65.851} {\bibfield
  {journal} {\bibinfo  {journal} {Rev. Mod. Phys.}\ }\textbf {\bibinfo {volume}
  {65}},\ \bibinfo {pages} {851} (\bibinfo {year} {1993})}\BibitemShut
  {NoStop}%
\bibitem [{\citenamefont {Lee}\ and\ \citenamefont {Yang}(1957)}]{LHY1}%
  \BibitemOpen
  \bibfield  {author} {\bibinfo {author} {\bibfnamefont {T.~D.}\ \bibnamefont
  {Lee}}\ and\ \bibinfo {author} {\bibfnamefont {C.~N.}\ \bibnamefont {Yang}},\
  }\href@noop {} {\bibfield  {journal} {\bibinfo  {journal} {Phys. Rev.}\
  }\textbf {\bibinfo {volume} {105}},\ \bibinfo {pages} {1119} (\bibinfo {year}
  {1957})}\BibitemShut {NoStop}%
\bibitem [{\citenamefont {Lee}\ \emph {et~al.}(1957)\citenamefont {Lee},
  \citenamefont {Huang},\ and\ \citenamefont {Yang}}]{LHY2}%
  \BibitemOpen
  \bibfield  {author} {\bibinfo {author} {\bibfnamefont {T.~D.}\ \bibnamefont
  {Lee}}, \bibinfo {author} {\bibfnamefont {K.}~\bibnamefont {Huang}}, \ and\
  \bibinfo {author} {\bibfnamefont {C.~N.}\ \bibnamefont {Yang}},\ }\href@noop
  {} {\bibfield  {journal} {\bibinfo  {journal} {Phys. Rev.}\ }\textbf
  {\bibinfo {volume} {106}},\ \bibinfo {pages} {1135} (\bibinfo {year}
  {1957})}\BibitemShut {NoStop}%
\bibitem [{\citenamefont {Lima}\ and\ \citenamefont
  {Pelster}(2011)}]{Pelster:PRA:2011}%
  \BibitemOpen
  \bibfield  {author} {\bibinfo {author} {\bibfnamefont {A.~R.~P.}\
  \bibnamefont {Lima}}\ and\ \bibinfo {author} {\bibfnamefont {A.}~\bibnamefont
  {Pelster}},\ }\href@noop {} {\bibfield  {journal} {\bibinfo  {journal} {Phys.
  Rev. A}\ }\textbf {\bibinfo {volume} {84}},\ \bibinfo {pages} {041604}
  (\bibinfo {year} {2011})}\BibitemShut {NoStop}%
\bibitem [{\citenamefont {Saito}(2016)}]{Saito:SocJap:2016}%
  \BibitemOpen
  \bibfield  {author} {\bibinfo {author} {\bibfnamefont {H.}~\bibnamefont
  {Saito}},\ }\href@noop {} {\bibfield  {journal} {\bibinfo  {journal} {Journal
  of the Physical Society of Japan}\ }\textbf {\bibinfo {volume} {85}},\
  \bibinfo {pages} {053001} (\bibinfo {year} {2016})}\BibitemShut {NoStop}%
\bibitem [{\citenamefont {Petrov}(2015)}]{Petrov}%
  \BibitemOpen
  \bibfield  {author} {\bibinfo {author} {\bibfnamefont {D.~S.}\ \bibnamefont
  {Petrov}},\ }\href {\doibase 10.1103/PhysRevLett.115.155302} {\bibfield
  {journal} {\bibinfo  {journal} {Phys. Rev. Lett.}\ }\textbf {\bibinfo
  {volume} {115}},\ \bibinfo {pages} {155302} (\bibinfo {year}
  {2015})}\BibitemShut {NoStop}%
\bibitem [{\citenamefont {Cabrera}\ \emph {et~al.}(2018)\citenamefont
  {Cabrera}, \citenamefont {Tanzi}, \citenamefont {Sanz}, \citenamefont
  {Naylor}, \citenamefont {Thomas}, \citenamefont {Cheiney},\ and\
  \citenamefont {Tarruell}}]{Tarruell:Science:2018}%
  \BibitemOpen
  \bibfield  {author} {\bibinfo {author} {\bibfnamefont {C.~R.}\ \bibnamefont
  {Cabrera}}, \bibinfo {author} {\bibfnamefont {L.}~\bibnamefont {Tanzi}},
  \bibinfo {author} {\bibfnamefont {J.}~\bibnamefont {Sanz}}, \bibinfo {author}
  {\bibfnamefont {B.}~\bibnamefont {Naylor}}, \bibinfo {author} {\bibfnamefont
  {P.}~\bibnamefont {Thomas}}, \bibinfo {author} {\bibfnamefont
  {P.}~\bibnamefont {Cheiney}}, \ and\ \bibinfo {author} {\bibfnamefont
  {L.}~\bibnamefont {Tarruell}},\ }\href@noop {} {\bibfield  {journal}
  {\bibinfo  {journal} {Science}\ }\textbf {\bibinfo {volume} {359}},\ \bibinfo
  {pages} {301} (\bibinfo {year} {2018})}\BibitemShut {NoStop}%
\bibitem [{\citenamefont {Baillie}\ \emph {et~al.}(2016)\citenamefont
  {Baillie}, \citenamefont {Wilson}, \citenamefont {Bisset},\ and\
  \citenamefont {Blakie}}]{Blakie:PRA:2016}%
  \BibitemOpen
  \bibfield  {author} {\bibinfo {author} {\bibfnamefont {D.}~\bibnamefont
  {Baillie}}, \bibinfo {author} {\bibfnamefont {R.~M.}\ \bibnamefont {Wilson}},
  \bibinfo {author} {\bibfnamefont {R.~N.}\ \bibnamefont {Bisset}}, \ and\
  \bibinfo {author} {\bibfnamefont {P.~B.}\ \bibnamefont {Blakie}},\
  }\href@noop {} {\bibfield  {journal} {\bibinfo  {journal} {Phys. Rev. A}\
  }\textbf {\bibinfo {volume} {94}},\ \bibinfo {pages} {021602} (\bibinfo
  {year} {2016})}\BibitemShut {NoStop}%
\bibitem [{\citenamefont {Cinti}\ \emph {et~al.}(2017)\citenamefont {Cinti},
  \citenamefont {Cappellaro}, \citenamefont {Salasnich},\ and\ \citenamefont
  {Macr\`{\i}}}]{Fabio:PRL:2017}%
  \BibitemOpen
  \bibfield  {author} {\bibinfo {author} {\bibfnamefont {F.}~\bibnamefont
  {Cinti}}, \bibinfo {author} {\bibfnamefont {A.}~\bibnamefont {Cappellaro}},
  \bibinfo {author} {\bibfnamefont {L.}~\bibnamefont {Salasnich}}, \ and\
  \bibinfo {author} {\bibfnamefont {T.}~\bibnamefont {Macr\`{\i}}},\
  }\href@noop {} {\bibfield  {journal} {\bibinfo  {journal} {Phys. Rev. Lett.}\
  }\textbf {\bibinfo {volume} {119}},\ \bibinfo {pages} {215302} (\bibinfo
  {year} {2017})}\BibitemShut {NoStop}%
\bibitem [{\citenamefont {Gross}(1958)}]{Gross:AnnPhys:1958}%
  \BibitemOpen
  \bibfield  {author} {\bibinfo {author} {\bibfnamefont {E.}~\bibnamefont
  {Gross}},\ }\href@noop {} {\bibfield  {journal} {\bibinfo  {journal} {Annals
  of Physics}\ }\textbf {\bibinfo {volume} {4}},\ \bibinfo {pages} {57 }
  (\bibinfo {year} {1958})}\BibitemShut {NoStop}%
\bibitem [{\citenamefont {Landau}(1937)}]{Landau:1937}%
  \BibitemOpen
  \bibfield  {author} {\bibinfo {author} {\bibfnamefont {L.}~\bibnamefont
  {Landau}},\ }\href@noop {} {\bibfield  {journal} {\bibinfo  {journal} {Zh.
  Eksp. Teor. Fiz.}\ }\textbf {\bibinfo {volume} {7}},\ \bibinfo {pages} {19}
  (\bibinfo {year} {1937})}\BibitemShut {NoStop}%
\bibitem [{\citenamefont {Macr\`{\i}}\ \emph {et~al.}(2013)\citenamefont
  {Macr\`{\i}}, \citenamefont {Maucher}, \citenamefont {Cinti},\ and\
  \citenamefont {Pohl}}]{Tommaso:PRA:2013}%
  \BibitemOpen
  \bibfield  {author} {\bibinfo {author} {\bibfnamefont {T.}~\bibnamefont
  {Macr\`{\i}}}, \bibinfo {author} {\bibfnamefont {F.}~\bibnamefont {Maucher}},
  \bibinfo {author} {\bibfnamefont {F.}~\bibnamefont {Cinti}}, \ and\ \bibinfo
  {author} {\bibfnamefont {T.}~\bibnamefont {Pohl}},\ }\href@noop {} {\bibfield
   {journal} {\bibinfo  {journal} {Phys. Rev. A}\ }\textbf {\bibinfo {volume}
  {87}},\ \bibinfo {pages} {061602} (\bibinfo {year} {2013})}\BibitemShut
  {NoStop}%
\bibitem [{\citenamefont {Cinti}\ \emph
  {et~al.}(2014{\natexlab{b}})\citenamefont {Cinti}, \citenamefont
  {Boninsegni},\ and\ \citenamefont {Pohl}}]{Cinti_2014}%
  \BibitemOpen
  \bibfield  {author} {\bibinfo {author} {\bibfnamefont {F.}~\bibnamefont
  {Cinti}}, \bibinfo {author} {\bibfnamefont {M.}~\bibnamefont {Boninsegni}}, \
  and\ \bibinfo {author} {\bibfnamefont {T.}~\bibnamefont {Pohl}},\ }\href
  {\doibase 10.1088/1367-2630/16/3/033038} {\bibfield  {journal} {\bibinfo
  {journal} {New Journal of Physics}\ }\textbf {\bibinfo {volume} {16}},\
  \bibinfo {pages} {033038} (\bibinfo {year} {2014}{\natexlab{b}})}\BibitemShut
  {NoStop}%
\bibitem [{\citenamefont {Gopalakrishnan}\ \emph {et~al.}(2010)\citenamefont
  {Gopalakrishnan}, \citenamefont {Lev},\ and\ \citenamefont
  {Goldbart}}]{cavityBEC}%
  \BibitemOpen
  \bibfield  {author} {\bibinfo {author} {\bibfnamefont {S.}~\bibnamefont
  {Gopalakrishnan}}, \bibinfo {author} {\bibfnamefont {B.~L.}\ \bibnamefont
  {Lev}}, \ and\ \bibinfo {author} {\bibfnamefont {P.~M.}\ \bibnamefont
  {Goldbart}},\ }\href {\doibase 10.1103/PhysRevA.82.043612} {\bibfield
  {journal} {\bibinfo  {journal} {Phys. Rev. A}\ }\textbf {\bibinfo {volume}
  {82}},\ \bibinfo {pages} {043612} (\bibinfo {year} {2010})}\BibitemShut
  {NoStop}%
\end{thebibliography}
\end{document}